\documentclass[letterpaper]{jpconf}
\usepackage{graphicx}
\usepackage{epsfig} 

\newcommand{\beq}{\begin{equation}}
\newcommand{\eeq}{\end{equation}}
\newcommand{\beqa}{\begin{eqnarray}}
\newcommand{\eeqa}{\end{eqnarray}}

\begin{document}
\title{Seeing Darkness: The New Cosmology} 

\author{Eric V.\ Linder}

\address{Berkeley Lab, University of California, Berkeley, CA 94720, USA} 

\ead{evlinder@lbl.gov}

\begin{abstract}
We present some useful ways to visualize the nature of dark energy and 
the effects of the accelerating expansion on cosmological quantities. 
Expansion probes such as Type Ia supernovae distances and growth probes 
such as weak gravitational lensing and the evolution of large scale 
structure provide powerful tests in complementarity.  We present a 
``ladder'' diagram, showing that in addition to 
dramatic improvements in precision, next generation probes will provide 
insight through an increasing ability to test assumptions of the 
cosmological framework, including gravity beyond general relativity. 
\end{abstract} 

\section{Introduction} 

{\it What happens when gravity is no longer an attractive force?\/}  
This provocative question is currently confronting cosmology.  What 
would be relegated to mere paradox a decade ago is now orthodox, 
thanks to burgeoning, precise observations of our universe.  The 
accumulated evidence from the Type Ia supernovae (SN) distance-redshift 
relation, cosmic microwave background radiation measurements, 
and large scale structure data have indicated ever more strongly that 
some 70\% of our universe acts in the paradoxical manner of increasing 
the acceleration of expansion of the universe under gravity. 

This is physics pointing to fundamentally new components or laws. 
One possibility is a quantum field with zeropoint energy filling all 
space, as in Einstein's cosmological constant $\Lambda$.  However we have 
no understanding of the magnitude of this energy density, nor why it 
should be dominating the cosmic dynamics today -- in the last factor 
of two in expansion out of the $10^{54}$ or so that have occurred in 
the history of the universe.  To consider other, possibly more 
tractable possibilities, we need to explore further frontiers in 
high energy physics, gravitation, and cosmology. 

The answer may lie in new physics of the quantum vacuum -- {\it Does 
nothing weigh something?\/}  Possibilities include dynamical scalar 
fields, or quintessence, or insights from string theory.  The answer 
may lie in new physics of gravitation -- {\it Is nowhere somewhere?\/} 
Possibilities include extra dimensions or other aspects of quantum gravity. 
Generically we refer to any accelerating 
physics as dark energy.  To guide us in the darkness we need new, 
highly precise data. 

In \S\ref{sec:obs} I briefly review the next generation probes that can 
provide direction to our quest for the nature of dark energy.  
\S\ref{sec:dia} presents a new method for visualizing the effects of 
acceleration and unifying aspects of cosmic expansion.  Some 
details of working with dynamical scalar fields are given in \S\ref{sec:sca}. 
Building a robust picture of the new cosmology through testing the 
physics framework and assumptions is emphasized in \S\ref{sec:ladder}. 

\section{Probing the Dark \label{sec:obs}} 

The geometric technique of measuring distances in the universe is 
currently the most direct, 
practical method of probing the acceleration of the universe.  
Observations giving luminosity in the case of Type Ia supernovae, or 
angular or radial scale in the case of baryon acoustic oscillation 
patterns in the large scale structure distribution, translate into 
cosmic distance and lookback time, and redshifts of the emitting 
objects give the 
scale factor at that time.  The dynamics of the expansion -- the 
changing in scale factor over time, $a(t)$ -- probes the acceleration. 

Other methods are more indirect, relying on secondary effects of 
the acceleration such as the slowing of mass growth seen through 
large scale structure measurements or the decay of gravitational potentials 
seen through cosmic microwave background (CMB) measurements.  For an 
introductory overview of using cosmological techniques to probe dark 
energy, see \cite{lindeal}. 

Next generation experiments are currently being designed specifically 
to shed light on dark energy.  In particular, they must be dedicated 
to controlling systematic uncertainties that would obscure or bias 
our view of dark energy.  One example is the Supernova/Acceleration 
Probe (SNAP; \cite{snap}), which will combine several of the leading 
techniques.  Supernovae distances will be traced back 10 billion years, 
over 70\% of the age of the universe, probing the redshift range $z=0-1.7$. 
A wide field survey optimized for weak gravitational lensing measurements 
gives complementary data, allowing the comparison of the expansion 
history and growth history emphasized in \S\ref{sec:ladder}.  Other 
less developed techniques such as baryon acoustic oscillations, cluster 
abundances, and strong lensing are enabled by the same data set.

\section{Visualizing the Dark \label{sec:dia}} 

Dark energy leads to acceleration of the expansion, but on a plot 
of cosmic scale factor vs.\ time, $a(t)$, the effect is fairly difficult 
to see by eye.  Moreover, we want to visualize not only {\it that\/} 
the expansion is accelerating but {\it how\/}, the subtle distinctions 
between models.  Here I present a new diagram (influenced by treatments 
of inflation -- early universe acceleration) where the effects are not 
only more obvious but intimately related to cosmological observables and 
theory. 

\begin{figure} 
\begin{center} 
\psfig{file=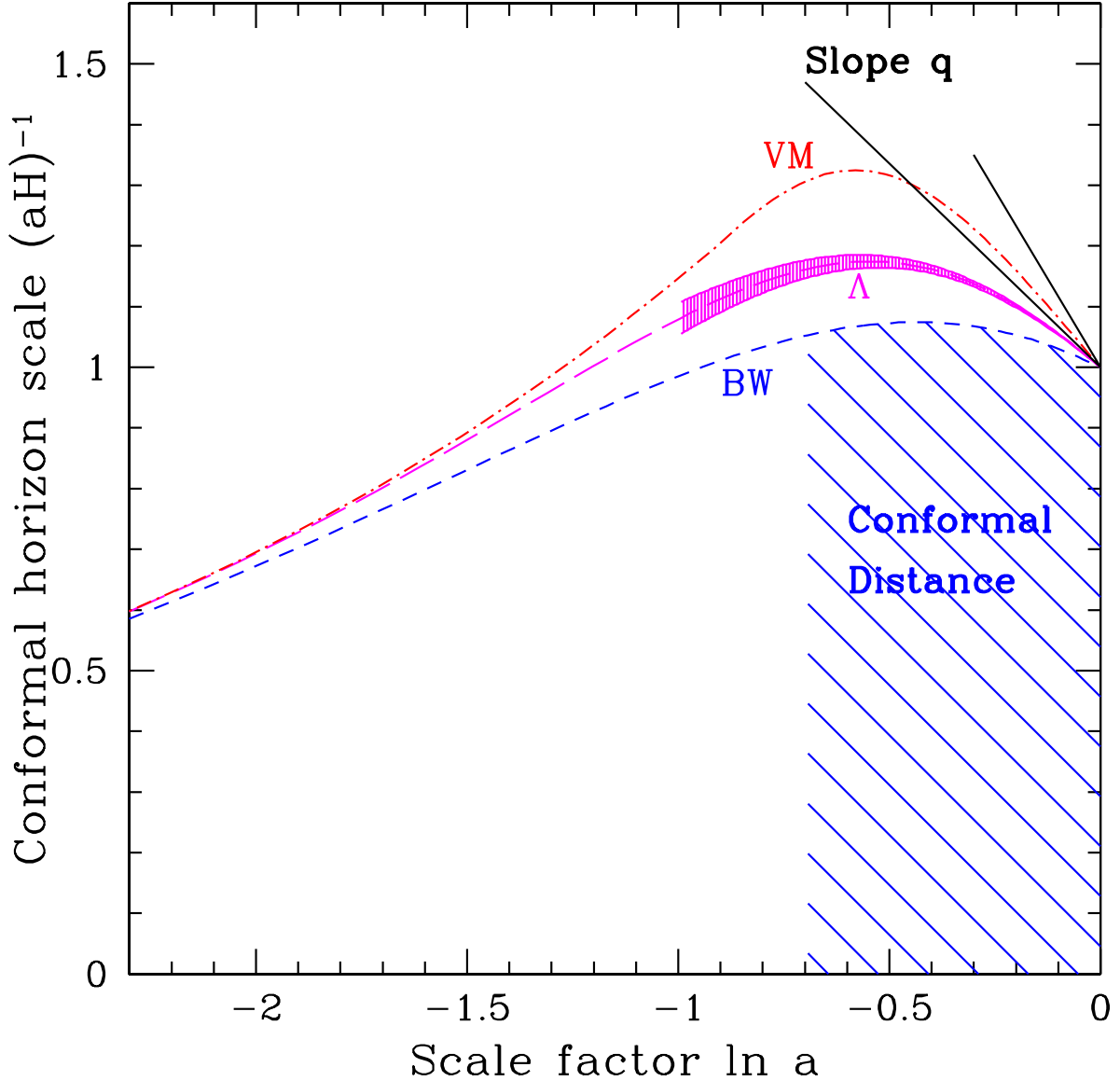,width=6.5in}
\caption{Visualization of dark energy cosmologies becomes intuitive 
in this diagram of the conformal Hubble scale vs.\ the logarithmic 
scale factor.  Curves are shown for three different models, all flat 
with matter density $\Omega_m=0.3$: a cosmological constant $\Lambda$, 
a braneworld cosmology (BW), and vacuum metamorphosis (VM).  The {\it area} 
under a curve between any two scale factors is the conformal distance, 
$\eta=d_{\rm lum}/(1+z)$, 
here shown by the diagonally shaded area for the braneworld model from 
today to $z=1$.  The {\it slope} of a curve indicates acceleration or 
deceleration, 
with the slope today being $q_0$.  This is shown for the $\Lambda$ curve 
by the long tangent, with $q_0=-0.55$.  The shorter line shows $q_0=-1$; 
slopes greater than this represent a superaccelerating expansion.  Note 
that while VM is a phantom ($w<-1$) model, it does not yet superaccelerate 
since the total equation of state is currently more positive than $-1$. 
The dense shading around the $\Lambda$ curve represents 95\% confidence 
level constraints possible with SNAP, clearly mapping both accelerating 
and decelerating phases and distinguishing dark energy physics. 
}
\label{fig:ahvisual}
\end{center}
\end{figure}

In such a diagram (a conformal horizon diagram), shown in 
Fig.~\ref{fig:ahvisual}, comoving wavelengths 
would simply be horizontal lines.  The slopes of the conformal Hubble 
horizon curves are $d(aH)^{-1}/d\ln a=q\,(aH)^{-1}$, or today 
simply $q_0$, the deceleration parameter.  
For an expanding universe, positive slopes therefore 
correspond to decelerating epochs, and a negative slope is the sign of 
acceleration.  

The area under a curve is simply the conformal distance 
$\eta=\int d\ln a (aH)^{-1}$, precisely the quantity that enters in 
luminosity distances and angular distances (up to redshift factors) in 
a flat universe.  Thus one can immediately see that distances in an 
accelerating universe are greater than they would be in a decelerating 
universe (with the same Hubble constant), or a less accelerating universe: 
compare the cosmological constant curve $\Lambda$ with the braneworld 
curve BW. 
One can also read off the total equation of 
state of the universe and its running: 
\beqa 
w_{\rm tot}&=&-\frac{1}{3}+\frac{2}{3}\frac{d\ln(aH)^{-1}}{d\ln a} \\ 
w'_{\rm tot}&=&\frac{2}{3}\frac{d^2\ln(aH)^{-1}}{d\ln a^2}. 
\eeqa

\section{Ghosts in the Dark \label{sec:sca}}

With the next generation data, such as the densely shaded region around 
the $\Lambda$ curve of Fig.~\ref{fig:ahvisual} representing the 95\% 
confidence limit of SNAP, we can hope to reveal the nature of 
dark energy.  As one example, we consider in this section the properties 
of a scalar field.  A canonical scalar field $\phi$ with Lagrangian 
${\cal L}=(1/2)\partial_\mu\phi \partial^\mu\phi - V(\phi)$ has an 
energy density $\rho=(1/2)\dot\phi^2+V$ and pressure $p=(1/2)\dot\phi^2-V$, 
where $V$ is the scalar field potential. 

General relativity instructs us that the gravitating mass is 
proportional to $\rho+3p$, so we can obtain an answer to our introductory 
question if the ratio $w\equiv p/\rho<-1/3$.  That is, for an equation 
of state ratio (EOS) $w<-1/3$, the field acts to accelerate the expansion. 
The quantity $w(z)$ plays a central role in understanding dark energy. 
In particular, both its value at any one time and its dynamics are 
important.  The standard approximation for taking both of these into 
account is the form 
\beq 
w(z)=w_0+w_a(1-a), \label{eq:wa} 
\eeq 
where the scale factor is related to the redshift through $a=(1+z)^{-1}$. 
This form has been shown to capture the essence of quintessence for 
a wide variety of models \cite{linprl}. 

Given the equation of state, one can formally obtain the energy density, 
the potential, the field dynamics, etc.  However in practice, this is 
nearly impossible, not merely because of observational uncertainties, 
but due to the intrinsic nature of our accelerating universe.  The 
field dynamics is given by 
\beq 
\dot\phi \sim \sqrt{(1+w)\rho} \sim HM_P\sqrt{1+w}, 
\eeq 
where $H=\dot a/a$ is the Hubble parameter and $M_P$ is the Planck energy. 
Since observations indicate that $1+w\ll1$, then the field has only 
rolled $\Delta\phi\sim \dot\phi/H\ll M_P$ during the epoch when dark 
energy significantly affects the universe.  Thus reconstruction of 
the potential is prevented. 

In a recent breakthrough, Caldwell \& Linder \cite{caldlin} realized 
that one can still 
use the field dynamics to categorize the nature of the physics responsible 
for the acceleration.  Measurements of the dynamics to a precision 
$\sigma(w'=\dot w/H)\approx 2(1+w)$ can distinguish between two distinct 
classes of ``freezing'' fields and ``thawing'' fields, with different 
physical origins.  The time variation $w'=-w_a/2$ in the parameterization 
of Eq.~(\ref{eq:wa}).  Note that without measurement of dynamics, only 
knowing an averaged value $\langle w\rangle$, say, the physics is almost 
completely obscured.  In particular, the entire ``thawing'' half of 
phase space would be mistaken for a cosmological constant. 

While the categorization strictly holds for canonical scalar fields, 
research in progress shows it to be more general.  But what about the 
possibility of $w<-1$, called phantom fields?  These are often shunned 
due to ``bad physics'', such as ghosts or imaginary mass particles and 
instabilities.  The consequences of $w<-1$ for our universe \cite{bigrip} 
are remarkably similar to the definition of ``bad'' given in the 1984 
movie {\it Ghostbusters\/}: ``imagine all life as you know it stopping 
instantaneously and every molecule in your body exploding at the speed 
of light''! 

However we must allow the data to lead us where it would.  Moreover, 
various ways of obtaining at least an effective $w<-1$ without ``bad'' 
physics have been theorized.  This ghostbusting includes 

\begin{itemize} 
\item Vacuum metamorphosis \cite{parkerraval} 
\item Coupled dark energy \cite{turner85,linder88,amendola,groexp} 
\item Curvature plus $\Lambda$ \cite{lincurv} 
\item Braneworld plus $\Lambda$ \cite{luestarkman,ckt05a} 
\item Climbing field \cite{ckt05} 
\end{itemize} 

Finally, any modification of the Friedmann expansion equation can 
be written in terms of an effective EOS \cite{lingrav}.  
Defining the deviation from the usual Friedmann equation, 
$\delta H^2\equiv (H/H_0)^2-\Omega_m a^{-3}$, simple 
examples include modifications depending solely on matter density, 
i.e.\ without any additional physical component in the universe: 

\beqa 
\delta H^2&=&A(\rho_m)^n \quad\Rightarrow\quad w=-1+n \\ 
\delta H^2&=&(8\pi/3)f(\rho_m) \quad\Rightarrow\quad 
w=-1+d\ln f/d\ln\rho_m 
\eeqa 
From any such effective $w(z)$ one can in turn construct an effective 
potential.

\section{The New Cosmology \label{sec:ladder}}

While the EOS describes the expansion history of the universe and 
contains critical clues to the fundamental physics, one still needs 
an underlying theory to specify the presence of microphysics such 
as field perturbations, or to distinguish between physical origins. 
That $w(z)$ is such a general language is a feature, but also a bug, 
preventing us from distinguishing two models with identical expansion 
histories (see \cite{groexp} for a detailed discussion).  Fortunately, 
the growth history of matter density perturbations gives another window 
on dark energy. 

Within general relativity, the perturbation growth is a function of the 
expansion history (plus perturbations in components other than matter, 
but these are expected to be negligible).  But more generally the 
growth also depends on the gravitation theory.  So we can distinguish 
the effects of dark energy as a physical component vs.\ those from 
a modification of gravity.  For a model independent approach, 
Linder \cite{groexp} proposed following the physics by incorporating the 
expansion effects via $w(z)$ and {\it separately\/} parameterizing 
the effects of gravity.  

The linear growth factor 
$g(a)=(\delta\rho_m/\rho_m)/a$ was found to be superbly approximated by 
\beq 
g(a)=e^{\int_0^a d\ln a\,[\Omega_m(a)^\gamma-1]}. \label{eq:ga} 
\eeq 
This is valid over a wide range of cosmologies to 0.05-0.2\%. 
The new parameter, the growth index $\gamma$, measures the modification 
of gravity: for scalar fields within general relativity, 
$\gamma=0.55+0.05[1+w(z=1)]$; for a braneworld model though 
$\gamma=0.68$. 

Beyond linearity, the growth of nonlinear matter structures such as 
galaxies and clusters 
of galaxies must be understood in dark energy cosmologies if they are 
to be used to probe dark energy.  This requires a large suite of 
cosmological simulations covering a wide range of parameter space, 
especially since current fitting formulas for the mass power spectrum, 
say, are accurate to only $\sim10\%$, and that primarily for 
cosmological constant universes.  However, \cite{nonlin} discovered 
a method for improving the accuracy by almost an order of magnitude, 
at the same time speeding up the parameter space coverage by $\sim100$. 
By matching the linear growth factor at two epochs, and $\Omega_m h^2$, 
they obtain accuracy better than 1.5\% in the {\it non\/}linear power 
spectrum. 
Their method automatically matches CMB constraints also.  This offers the 
hope of rapid development in the accurate calculation of quantities depending 
on the power spectrum (such as weak gravitational lensing measurements 
of large scale structure mass growth) over a variety of dark energy models. 

Thus, cosmological measurements giving the expansion history (e.g.\ 
through supernovae) and the growth history (e.g.\ through weak gravitational 
lensing), working together, promise real hope to reveal the origin of dark 
energy.  A next generation dark energy experiment must include both 
approaches to understand the nature of the new physics. 

\begin{figure}[h] 
\begin{center}
\psfig{file=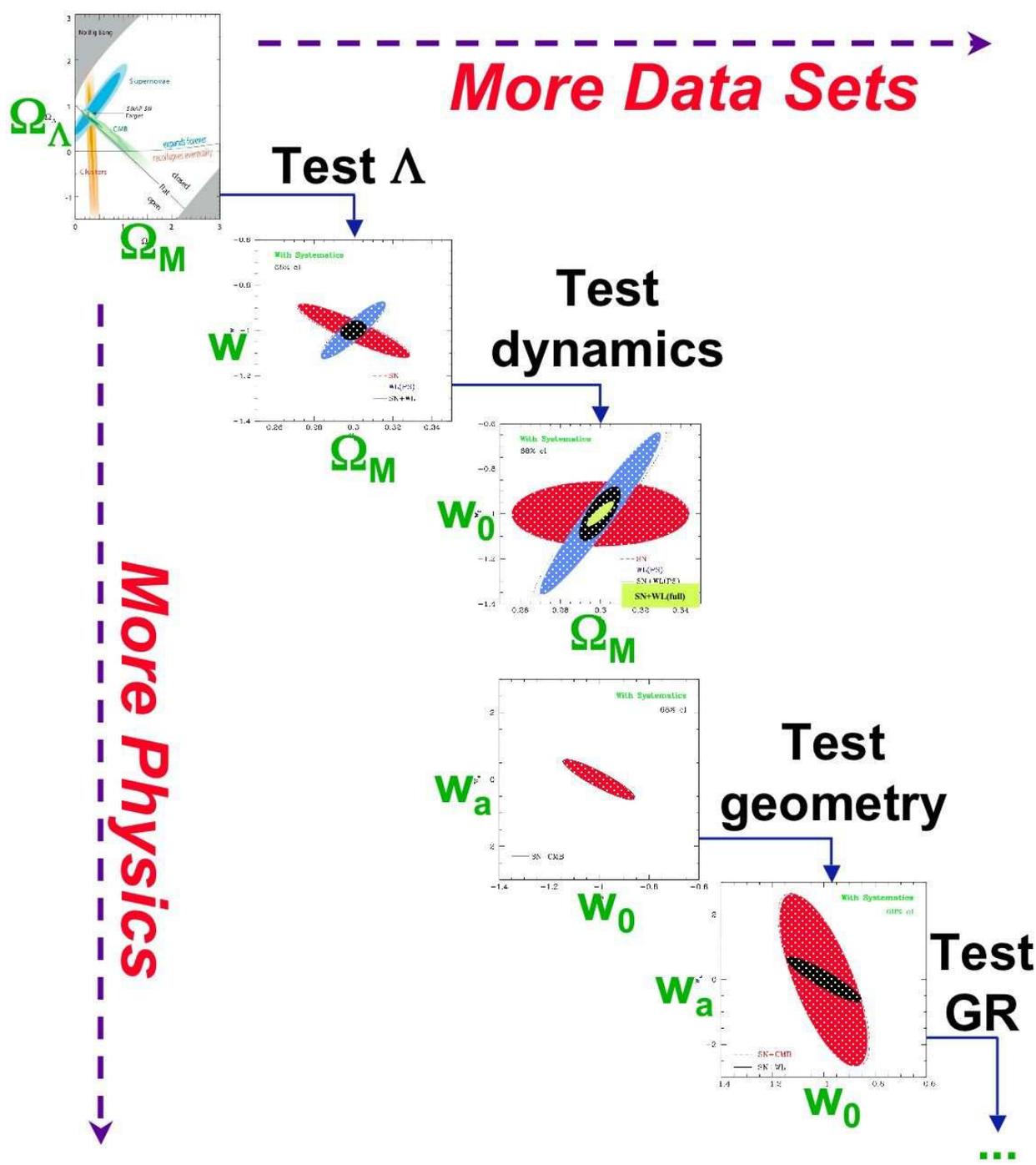,width=6.5in}
\caption{Assumptions imposed on dark energy physics allow strong 
constraints from data, but are often unjustified given our ignorance 
of dark energy.  Relaxing these priors sometimes dramatically increase 
the uncertainties -- as when avoiding assumption that the equation of 
state is constant -- but allows us to test for important physics.  The 
addition of robust, complementary, systematics controlled data sets, 
again constrains the nature of dark energy.  (This diagram is intended 
to be illustrative only, not quantitative, but does 
include systematics.)  This ``ladder'' of 
advancing the physics and the data hand in hand represents the new 
cosmology of seeing darkness. 
}
\label{fig:ladder} 
\end{center}
\end{figure}

We illustrate this in Fig.~\ref{fig:ladder}.  If we ``weigh'' dark energy 
in a diagram of dimensionless vacuum energy density vs.\ matter density, 
we need to remember that this assumes a cosmological constant is the 
dark energy; if we merely {\it allow\/} for a ``springiness'' of space 
$w\ne-1$, then the parameter estimation contours will increase.  However 
we have no evidence for or expectation that $w$ is constant: time variation 
is generic, the ``stretchiness'' of the spring.  Merely allowing for this 
possibility blows up the confidence region and we must employ additional 
data sets to constrain it.  This is 
the power of complementarity -- it allows us to be more physically 
reasonable, not to impose a priori constraints but let the data guide 
the way.  

We can pinpoint the physics through measuring the dynamics $w_a$.  Suppose 
then we allow for the presence of spatial curvature; again the constraints 
weaken and again we need to bring in complementary data.  Suppose then 
we allow for modification of gravity, e.g.\ in the model independent 
parameter space of $w_0$, $w_a$, $\gamma$\dots.  

We refer to this process of relaxing the assumptions imposed on the physics 
and then combining additional probes as the ladder of constraints. 
To achieve true 
understanding of the nature of our universe, we require next generation 
experiments utilizing all the robust cosmological techniques we have. 
It will be an exciting decade as we learn to see darkness.

\ack 
I thank the organizers of TAUP2005 for the invitation and hospitality.  
I gratefully 
acknowledge Gary Bernstein, Dragan Huterer, and Masahiro Takada for 
information enabling Fig.~\ref{fig:ladder}.  This work was 
supported in part by the Director, Office of Science,
US Department of Energy under grant DE-AC02-05CH11231.

\end{document}